# Growth mechanism of nanostructured superparamagnetic rods obtained by electrostatic co-assembly

M. Yan[a], J. Fresnais[b] and J.-F. Berret[a]*

[a]Matière et Systèmes Complexes, UMR 7057 CNRS Université Denis Diderot Paris-VII, Bâtiment Condorcet, 10 rue Alice Domon et Léonie Duquet, 75205 Paris (France)
[b]UPMC Paris VI – Laboratoire de Physico-chimie des Electrolytes, Colloïdes et Sciences Analytiques UMR 7195 CNRS 4 place Jussieu, 75252 Paris Cedex 05 (France)

**Abstract**
We report on the growth of nanostructured rods fabricated by electrostatic co-assembly between iron oxide nanoparticles and polymers. The nanoparticles put under scrutiny, $\gamma$-$Fe_2O_3$ or maghemite, have diameter of 6.7 nm and 8.3 nm and narrow polydispersity. The co-assembly is driven by *i)* the electrostatic interactions between the polymers and the particles, and by *ii)* the presence of an externally applied magnetic field. The rods are characterized by large anisotropy factors, with diameter ~ 200 nm and length comprised between 1 and 100 µm. In the present work, we provide for the first time the morphology diagram for the rods as a function of ionic strength and concentration. We show the existence of a critical nanoparticle concentration ($c_c = 10^{-3}$ wt. %) and of a critical ionic strength ($I_S^c = 0.42$ M) beyond which the rods do not form. In the intermediate regimes ($c = 10^{-3} - 0.1$ wt. % or $I_S = 0.35 - 0.42$ M), only tortuous and branched aggregates are detected. At higher concentrations and lower ionic strengths, linear and stiff rods with superparamagnetic properties are produced. Based on these data, a mechanism for the rod formation is proposed. The mechanism proceeds in two steps : *i)* the formation and growth of spherical clusters of particles, and *ii)* the alignment of the clusters induced by the magnetic dipolar interactions. As far as the kinetics of these processes is concerned, the clusters growth and their alignment occur concomitantly, leading to a continuous accretion of particles or small clusters, and a welding of the rodlike structure.

*Corresponding author : jean-francois.berret@univ-paris-diderot.fr

## 1 - Introduction
In the past decade, large research efforts were directed towards the fabrication of new nanostructured materials using inorganic nanoparticle (NP) as elementary building blocks. To this end, supracolloidal assemblies such as spherical clusters, elongated rods and wires and planar sheets were designed for various applications, including photonics,[1-3] electronics,[1, 4, 5] sensing,[6-9] and biomedicine.[10-12] In this fast-evolving field, one-dimensional structures specifically made from magnetic particles have also attracted much attention, essentially because of the possibility to build large assemblies with purely superparamagnetic behaviors.[13-17] It is indeed well-known that the magnetic properties of individual nanocrystals depend strongly on their size, and that for iron-based particles such as maghemite ($\gamma$-$Fe_2O_3$) or magnetite ($Fe_3O_4$), the onset of ferromagnetism and multidomain structure are classically observed for particles with sizes larger than 30 nm.[18] In most instances reported in the literature, elongated magnetic structures were made from commercially available microbeads of size ~ 1 µm.[19, 20, 13, 7, 14-16, 21-23] The magnetic colloids utilized were in general polystyrene beads loaded with superparamagnetic 10 nm NPs that were chemically embedded in the polymer during the bead synthesis. In these approaches, permanent chains were formed in a two-step process that consisted first in the application of a magnetic field for the alignment of the beads thanks to the magnetic dipolar interactions, and second in a physical or chemical linking of the beads via polymers or ligands. As a result, filaments and chains with diameter ~ 1 µm and length 1 – 200 µm were fabricated. These filaments were either dispersed in a solvent[19, 13-16] or tethered on substrates,[7, 16, 17] and depending on the linking agents exhibited various degrees of flexibility and mechanical properties. Furst and Gast measured the rupture tensions of linear assembly of polystyrene particles functionalized by streptavidin and biotin linkers and found forces in the pico Newton range.[13] Goubault *et al.* investigated the hairpin configurations of such filaments submitted to increasing magnetic field, and from the curvature-field intensity dependence derived the bending rigidity and Young modulus of the arrays, which was found around 1 kPa.[7] Permanent and elongated arrays of particles of this kind have been designed to serve as micromechanical tools, such as actuators, sensors or mixers for microfluidics devices.[14, 16, 21, 17, 23]

In contrast to the large body of work on microbeads forming-filaments, very few attempts were made to build permanent structures comprising individual 10 nm particles.[24-26, 10, 27, 28, 12] Few strategies exist however and they generally utilized neutral or ion-containing polymers which then served as a template during the nanocrystal synthesis,[26, 10, 12] or as oppositely charged species for electrostatic complexation schemes.[25, 27, 28] Sheparovych *et al.* used a magnetic field for the fabrication of chain-like structures deposited *in-situ* on a substrate and resulting from the interactions between anionically charged $Fe_3O_4$ particles and a cationic polyelectrolyte.[25] Tortuous filaments of iron oxide nanocrystals were produced by the reaction of Fe(II) and Fe(III) salts in presence of dextran and were tested positively for *in vivo* targeting and imaging in a mouse model.[12] These authors also showed that nanomaterials that are elongated along one dimension were able to avoid the reticuloendothelial system of the mouse more efficiently than spherical colloids with the same volume. For sake of completeness, it should be mentioned that alternative approaches based on epitaxial growth or templates were also performed for the design of ferromagnetic rods, *i.e.* rods displaying permanent magnetic moments even without magnetic field.[29-31]

Here, we report on nanostructured rods fabricated by electrostatic co-assembly between iron oxide nanoparticles and polymers.[27, 32] The rods (diameter 200 nm, length 1 – 100 µm) were obtained by dialyzing salted dispersions containing anionic maghemite NPs and cationic copolymers under the application of a magnetic field. Transmission electron microscopy showed that inside the rods the NPs were linked together by the cationic polymer "glue" and remained so even after removing the external magnetic field.[27] In the present work, we investigate the mechanism of formation of the rods by varying physical and physico-chemical parameters such as the particle size, the concentration and the ionic strength of the dispersions. It is found that this mechanism proceeds in two steps : the formation of spherical clusters of particles and the alignment of the clusters induced by the magnetic field, the two processes occurring simultaneously.

# 2 – Materials and Methods

## 2.1 – Chemicals and nanoparticles

The anionically charged NPs have been co-assembled with a cationic–neutral diblock copolymers, referred to as poly (trimethylammonium ethylacrylate)-*b*-poly (acrylamide). The copolymers were synthesized by MADIX® controlled radical polymerization, which is a Rhodia patented process.[33, 34] Light scattering experiment was performed on the copolymer aqueous solutions to determine the weight-averaged molecular weight $M_W$ (= 44400 ± 2000 g mol$^{-1}$) and mean hydrodynamic diameter $D_H$ (= 11 nm) of the chains.[35] The molecular weights targeted by the synthesis were 11000-*b*-30000 g mol$^{-1}$, corresponding to 41 monomers of trimethylammonium ethylacrylate methylsulfate and 420 monomers of acrylamide, in fair agreement with the experimental values. In the following, this polymer will be abbreviated as PTEA$_{11K}$-*b*-PAM$_{30K}$.[35] The polydispersity index was determined by size exclusion chromatography at 1.6.

| $\gamma$-Fe$_2$O$_3$ – nanoparticles | NP6.7 | NP8.3 | technique |
|---|---|---|---|
| $D_0^{VSM}$ (nm) | 6.7 ± 0.2 | 8.3 ± 0.2 | VSM |
| $s^{VSM}$ | 0.21 ± 0.03 | 0.21 ± 0.03 | VSM |
| $D_0^{TEM}$ (nm) | 6.8 ± 0.2 | 9.3 ± 0.2 | TEM |
| $s^{TEM}$ | 0.21 ± 0.02 | 0.18 ± 0.02 | TEM |
| $D_H$ (nm) – bare | 17 ± 1 | 20 ± 1 | DLS |
| $D_H$ (nm) – coated | 22 ± 1 | 25 ± 1 | DLS |

**Table I** : *Median diameter ($D_0$), polydispersity (s), and hydrodynamic diameter $D_H$ of the bare and PAA$_{2K}$-coated NPs studied in this work. The particle size distributions were found to be log-normal for the two batches (Eq. SI-1). $D_0$ and s were measured using vibrating sample magnetometry (VSM) and transmission electron microscopy (TEM).*

The synthesis of the superparamagnetic NPs investigated here was elaborated by Massart using the technique of « soft chemistry ».[36] Based on the polycondensation of metallic salts in alkaline aqueous media, this technique resulted in the formation of magnetite (Fe$_3$O$_4$) NPs of sizes comprised between 4 and 15 nm. Magnetite crystallites were further oxidized into maghemite ($\gamma$-Fe$_2$O$_3$) and sorted according to their size. In the conditions of the synthesis (pH 1.8, weight concentration c ~ 10 wt. %), the magnetic dispersions were stabilized by electrostatic interactions arising from the native cationic charges at the surface of the particles. In this work, two batches of particles of diameters 6.7 and 8.3 nm, hereafter noted NP6.7 and NP8.3 were synthesized. The particle size distributions were characterized by Vibrating Sample Magnetometry (VSM), Transmission Electron Microscopy (TEM) and Dynamic Light Scattering (DLS) (see Supporting Information, SI-1). In Table I are listed the values of the diameters $D_0^{VSM}$, $D_0^{TEM}$, $D_H$ and polydispersities obtained by these methods. Note the good agreement between the VSM and TEM data, and the systematic shift of the hydrodynamic diameters $D_H$ with respect to the bare values. This reasons of the shift between the geometric and hydrodynamic dimensions are well-known : *i)* the particles are slightly anisotropic (aspect ratio 0.8); *ii)* when particles are distributed as this is the case here, light scattering is more sensitive to the largest particles of the distribution. In order to improve their colloidal stability, the cationic particles were further coated by poly(acrylic acid) oligomers with molecular weight 2000 g mol$^{-1}$, using the precipitation–redispersion process described previously.[37] The hydrodynamic sizes found in γ-Fe$_2$O$_3$-PAA$_{2K}$ dispersions were 5 nm above that of the bare particles, indicating the presence of a 2.5 nm PAA$_{2K}$ brush surrounding the particles.

**2.2 - Fabrication of the nanostructured rods**

Fig. 1 describes the protocol that controlled the NPs co-assembly in the presence of a magnetic field. The adopted strategy involved in a first step the preparation of two separate 1 M NH$_4$Cl solutions containing respectively *i)* the anionic iron oxide NPs (γ-Fe$_2$O$_3$) and *ii)* the PTEA$_{11K}$-*b*-PAM$_{30K}$ diblock copolymers. The γ-Fe$_2$O$_3$ concentrations put under scrutiny here ranged between $10^{-4}$ to $3\times10^{-1}$ wt. %, with a polymer-to-nanoparticle volume ratio of 0.5. In a second step, the two solutions were mixed with each other and it was checked by dynamic light scattering that the two components remained dispersed ($D_H = 21 \pm 1$ nm).[27]

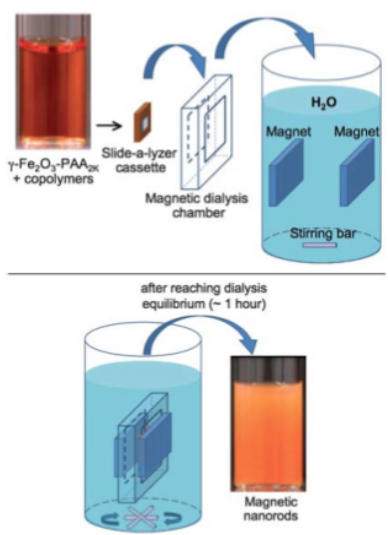

*Figure 1 :* *Schematic representation of the protocol that controlled the nanoparticle co-assembly into superparamagnetic nanostructured rods in the presence of external magnetic field.*

In a third step, the ionic strength of the mixture was progressively diminished by dialysis in the presence of an external magnetic field (0.3 T). Dialysis was performed against de-ionized water using a Slide-a-Lyzer® cassette with MWCO of 10 kD (Thermo Scientific). The volume of the dialysis bath was 300 times larger than that of the samples. The electrical conductivity of the dialysis bath was measured during the ion exchange and served to monitor the desalting kinetics. In the condition described here, the whole process reached a stationary and final state within 50 – 100 mn. Once the ionic strength of the bath reached its stationary value, typically $10^{-3}$ M, the external magnetic field was removed and the dispersions were studied by optical microscopy. The dialysis experiment between the initial and final ionic strengths was characterized by an average rate of ionic strength change $dI_S/dt \sim -10^{-4}$ M s$^{-1}$. Note that with dialysis, the iron oxide concentration remained practically constant.

**2.3 - Experimental Techniques**

For optical microscopy, phase-contrast images of the rods were acquired on an IX71 inverted microscope (Olympus) equipped with 40× and 60× objectives. 5 µl of dispersion at concentration 0.1 wt. % were deposited on a glass plate and sealed into to a Gene Frame® (Abgene/Advanced Biotech) dual adhesive system. We used a Photometrics Cascade camera (Roper Scientific) and Metaview software (Universal Imaging Inc.) as acquisition system. In order to determine the length distribution of the rods, pictures were digitized and treated by the ImageJ software (http://rsbweb.nih.gov/ij/). Transmission Electron Microscopy (TEM) was carried out on a Jeol-100 CX microscope at the SIARE facility of University Pierre et Marie Curie (Paris 6). TEM was used to both characterize the individual γ-Fe$_2$O$_3$ NPs (magnification 160000×) and the superparamagnetic nanostructured rods (magnification from 10000× to 100000×). Electrical conductivity measurements during dialysis were monitored by a Cyberscan PC6000 operating with a 4-cell conductivity electrode.

# 3 - Results and Discussion

**3.1 - Superparamagnetic nanostructured rods**

Fig. 2a shows an optical transmission microscopy image of rods made of 6.7 nm γ-Fe$_2$O$_3$ particles at concentration c = 0.1 wt. % and polymer-to-nanoparticle volume ratio 0.5. In the absence of magnetic field, anisotropic structures with random orientations with typical lengths lying between 1 and 100 µm are clearly visible. Series of images similar to that of Fig. 2a were analyzed quantitatively to retrieve the rod length distribution for the two samples investigated in this study, NP6.7 and NP8.3. In both cases, the length distribution was found to be well accounted for by a log-normal function of the form:

$$p(L, L_0, s_L) = \frac{1}{\sqrt{2\pi}\beta_L(s_L)L} \exp\left(-\frac{\ln^2(L/L_0)}{2\beta_L(s_L)^2}\right) \qquad (1)$$

where $L_0$ is defined as the median length and $\beta_L(s_L)$ is related to the polydispersity index $s_L$ by the relationship $\beta_L(s) = \sqrt{\ln(1+s_L^2)}$. The polydispersity index is defined as the ratio between the standard deviation $(\langle L^2 \rangle - \langle L \rangle^2)^{1/2}$ and the average length $\langle L \rangle$. Note that the same type of distribution was assumed for the diameters of the NPs derived TEM. For rods made from NP6.7 and NP8.3, one obtained $L_0$ = 8.2 ± 0.7 and 17.5 ± 0.9 µm respectively. The polydispersity $s_L$ was similar for the two specimens and equal to 0.5 (see Supporting Information, SI-2). If a magnet was brought near to the sample, the rods reoriented spontaneously and followed the magnetic field lines (Fig. 2b). The coupling between the rods and the external field was shown to originate from the superparamagnetic properties exhibited by the elongated structures.[27] Even after a prolonged period under the application of a field (B = 12 mT), the rods remained dispersed and unaggregated. As an illustration of these orientational behavior, a movie of rods subjected to a rotating magnetic field (at the frequency of 0.2 Hz) is shown in Supporting Information (Movie#1). These properties could be exploited in microfluidics as well as in therapeutics for the elaboration of micro-actuators.[11]

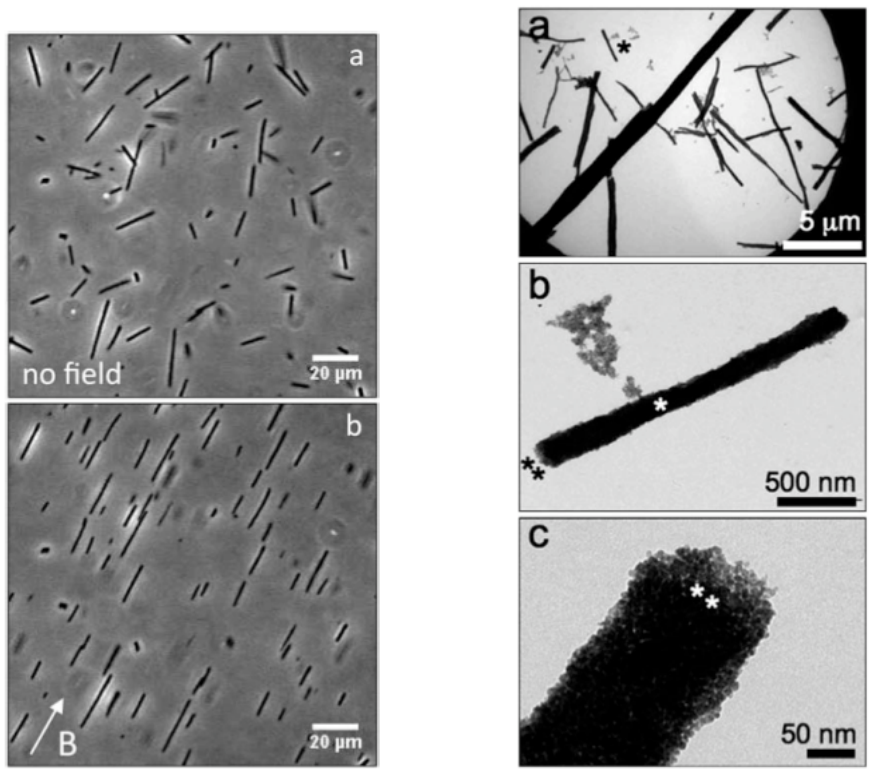

*Figure 2 :* Phase-contrast optical microscopy images (40×) of a dispersion of nanostructured rods made from 6.7 nm γ-$Fe_2O_3$ particles. In absence of magnetic field (a), the rods are randomly oriented. With a magnetic field of 10 mT applied in the plane of observation (b), the rods align along with the field.

*Figure 3 :* TEM images of nanostructured rods constituted of γ-$Fe_2O_3$ nanoparticles and obtained at magnifications 2600× (a), 26000× (b) and 160000× (c). The upper panel (a) displays rods of different lengths. The intermediate panel (b) shows one rod of length of 2.25 μm and diameter of 200 nm (indicated by a star). In the lower panel (c), the rod extremity was zoomed so as to display the individual particles tightly held together and forming the core cylindrical structure (indicated by two stars).

In order to gain insight into the microscopic structure of the rods, TEM was performed on dilute dispersions (c = 0.01 wt. %). Figs. 3a, 3b and 3c show images of rods at magnification 2600×, 26000× and 160000×, respectively. In Fig. 3a, rods of various diameters (200 nm – 1 μm) and lengths (1 – 20 μm) were observed. For the thickest rods, as that spreading along the diagonal of the image in Fig. 3a, it can be seen that they resulted from the parallel association of thinner rods. At this point, it is however unclear if the parallel clustering of the rods was achieved during the dialysis process, or at the evaporation of the solvent on the TEM grid. We here focus on the elementary anisotropic colloids, as the one indicated by a star. Fig. 3b emphasizes such an example of rod displaying a regular cylindrical body. For this rod, the length is 2.25 µm and the diameter 200 nm. In Fig. 3c, the extremity of the rod was enlarged so as to display the individual particles tightly held together and forming the cylindrical core structure (indicated by two stars). In a recent publication, the volume fraction occupied by the nanoparticles in such aggregates was assessed by small-angle neutron scattering to be 25 % of the whole, corresponding to $5\times10^4$ NPs per micrometer of length.[27]

### 3.2 – Rod growth as a function of the dialysis time

The dialysis of the dispersion containing the NPs and the polymers was performed over time periods comprised between 1 to 100 mn. Electric conductivity measurement of the dialysis bath allowed us to monitor the time evolution of the ionic strength outside the cassette, this conductivity increasing from $\Lambda = 2.5$ µS m$^{-1}$ at the initial ionic strength ($I_S \sim 0$ M) to $\Lambda = 60$ µS m$^{-1}$. This later conductivity corresponded to an ionic strength $I_S \sim 10^{-3}$ M, estimated from the volume of the cassette with respect to that of the water bath. From conductivity measurement, it was shown that the time necessary to reach the stationary state was estimated to be typically 60 mn. The ionic strength $I_S$ of the NP dispersion inside the cassette was then calculated and plotted as a function of time in Fig. 4a (sample NP6.7, c = 0.1 wt. %). The desalting kinetics could be adjusted by a single exponential function of the form $I_S(t) = I_S^0 \exp(-t/\tau)$, where $I_S^0$ denotes the initial ionic strength ($I_S^0 = 1$ M) and $\tau$ the relaxation time ($\tau = 12.5 \pm 0.2$ mn). The behavior of the electrical conductivity was found to be highly reproducible from one dialysis to another, as well as from one dialysis cassette (with the same cutoff) to another.

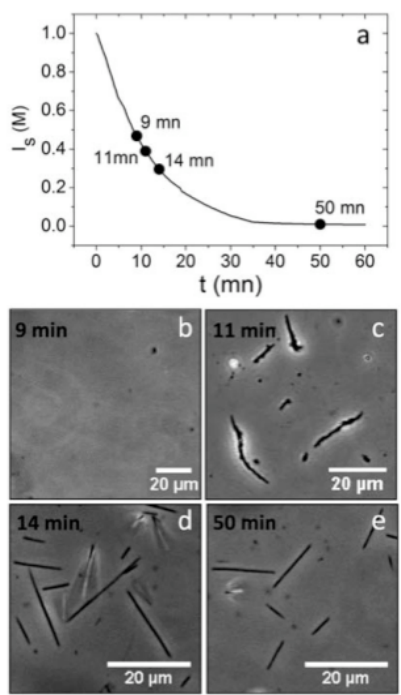

*Figure 4* : The ionic strength $I_S$ of the nanoparticle dispersion (sample NP6.7, c = 0.1 wt. %) inside the cassette is shown as a function of dialysis time (a). Phase-contrast optical microscopy images (40×) of the NP/polymer dispersions (sample NP6.7, c = 0.1 wt. %) at time t = 9 mn (b), 11 mn (c), 14 mn (d) and 50 mn (e), corresponding to ionic strengths $I_S$ = 0.45, 0.38, 0.29 and 0.001 M.

Successive dialysis experiments were performed and stopped at different dialysis times up to 50 mn, the dispersions being rapidly removed from the cassette and examined by optical microscopy. In Figs. 4b-4e are displayed 4 images of the γ-Fe$_2$O$_3$/PTEA$_{11K}$-*b*-PAM$_{30K}$ dispersions at time t = 9, 11, 14 and 50 mn, corresponding to ionic strength I$_S$ = 0.45, 0.38, 0.29 and 0.001 M. Note that at the transition the dialysis was characterized by an average rate of ionic strength change dI$_S$/dt ~ -5×10$^{-4}$ M s$^{-1}$.[27, 38] In a first period, comprised between the starting of the dialysis and *t* = 9 mn, no microscopic aggregates could be detected (Fig. 4b). Two minutes later (t = 11 mn), elongated and tortuous structures, hereafter dubbed wire-like aggregates with contour length of a few microns showed up (Fig. 4c). As time evolved (t = 14 and 50 mn), linear and stiff nanostructured rods were finally observed (Figs. 4d and 4e). It was checked that dialysis lasting from 60 and 1000 mn yielded rods with the same length distribution. The study on the NP8.3 sample revealed similar ionic strength variation and growth processes (see data in Supporting Information).

Figs. 5a and 5b describe the time and ionic strength evolution of the average rods length for the two samples. Here, three ionic strength or time regimes can be distinguished, hereafter noted I, II and III. The onset of aggregation corresponding to the transition between regime I and regime II was found at the critical ionic strength I$_S^c$ = 0.42 M. In regime II, the aggregates were wire-like, exhibiting tortuous structures and few lateral branching. There, the average length increased rapidly up to their stationary values, at L$_0$ = 8.2 and 17.5 μm respectively. Regime III found for t > 15 mn and I$_S$ < 0.3 M was characterized by the formation of a unique morphology for the rods and a saturation of their median length. Note that in all cases, the length distribution remained large and constant (s$_L$ = 0.5).

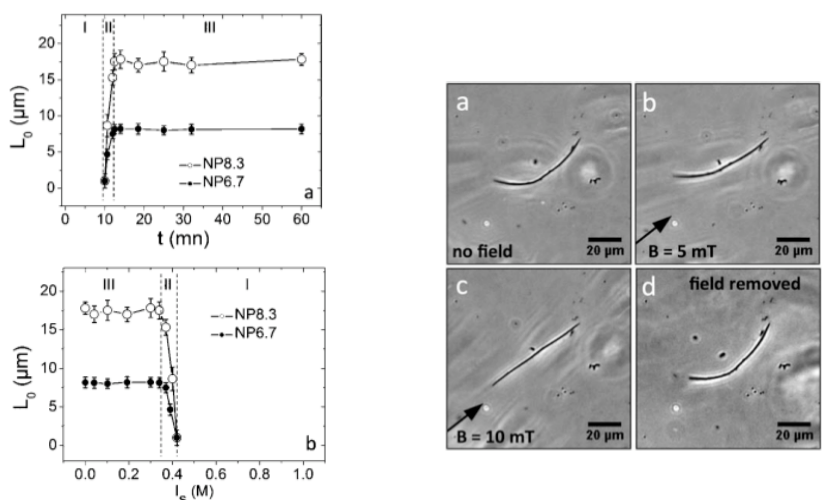

***Figure 5*** *: Time (a) and ionic strength (b) evolutions of the average rod length obtained with the NP6.7 and NP8.3 dispersions. Three morphology regimes including dispersed particles (Regime I), wire-like aggregates (Regime II) and rigid rods (Regime III) were identified for the nanoparticle co-assembly. For the wire-like aggregates, the length was estimated as the contour length of the objects.*

***Figure 6*** *: Phase-contrast microscopy images of a loose and flexible wire-like aggregate obtained in Regime II (0.35 M < I$_S$ < 0.42 M). In (b) and (c), the aggregate was submitted to a*

*magnetic field B = 5 mT and 10 mT respectively, yielding a stretching of the structure. After removal of the field (d), the stretched aggregate returned to its initial conformation.*

In the time sequence depicted in Fig. 5, it may seem surprising that wire-like aggregates with irregular morphologies of regime II could be the precursors of the linear and rigid rods found in regime III. In order to investigate this issue, an external magnetic field of varying magnitude (B = 0 – 12 mT) was applied to the dispersion shown in Fig. 4c. Figs. 6 provide optical microscopy images of such an aggregate, and show that the aggregate could be stretched by the application of a magnetic field (Fig. 6b and 6c). After removing the field, the stretched structure returned to its initial wire-like state (Fig. 6d). A movie of this transition was displayed in Supporting Information (Movie#2). This result indicates that the wire-like structures in regime II were loose and flexible. This property was attributed to the screened electrostatic interactions between particles and polymers. With decreasing $I_S$, the electrostatic interactions strengthened and the particles were frozen in the state of elongated and rigid rods. With this additional experiment, it can be understood that the precursor state of rigid rods can be wire-like and tortuous in absence of applied magnetic field.

### 3.3 – Rod growth as a function of the concentration

For the study of the role of the NP concentration c, the dialysis time was fixed at 60 mn, corresponding to an ionic strength decrease of ∼ 1 M, as illustrated in Fig. 4a. The γ-$Fe_2O_3$ concentrations put under scrutiny ranged between $10^{-4}$ to $3×10^{-1}$ wt. %, with a polymer-to-nanoparticle ratio of 0.5. For the most diluted samples, typically below $5×10^{-2}$ wt. %, the dispersions obtained after the dialysis process were concentrated by magnetic sedimentation or centrifugation and studied by optical microscopy. As for the kinetic study, the concentration diagram could be divided in three distinct regimes noted I, II and III. At very low concentration, below the critical concentration $c_c = 10^{-3}$ wt. % (regime I), no micronic aggregates could be observed. In regime II, for $c_c < c < c^* = 10^{-2}$ wt. %, wire-like aggregates with median contour length of a few microns were observed. Their morphologies were very similar to those obtained by varying the ionic strength. Above a transition concentration $c^* = 10^{-2}$ wt. % (regime III), the morphology became that of linear and rigid rods with lengths comprised between 1 and 100 μm.

Figs. 7 display images of the elongated structures found in regime II (Fig. 7a and 7b) and in regime III (Fig. 7c and 7d). The analysis of the rod lengths yielded distributions in agreement with Eq. 1 at all concentrations. The median lengths $L_0$ were found to increase rapidly with increasing concentration, up to c* where it reached a pseudo-plateau (Fig. 8). Note that over the whole c-range, the NP8.3 rods were always above the rods achieved with the NP6.3, indicating a strong dependence of the growth kinetics with respect to the initial particle diameter. Note also that for the two NPs, the critical c* remained identical. Both ionic strength and concentration dependences were summarized in a general morphology diagram valid for magnetic rods obtained from this desalting method (Fig. 9). The figure highlights the three regimes already mentioned : dispersed particles or clusters (Regime I), wire-like aggregates (Regime II) and rigid rods (Regime III).

In order to evaluate the potential conformational changes of the wire-like aggregates, an external magnetic field of varying magnitude was applied to the samples of regime II ($6×10^{-3}$ wt.%). The procedure used was identical to that of Section 3.2. The field was progressively increased from B = 0 to 12 mT and the wire-like conformation remained unchanged. Within the optical microscopy accuracy (~ 0.5 µm), the aggregates could not be stretched nor elongated by the application of a field. A movie in the Supporting Information (Movie#3) section shows this behavior. This result confirmed that at concentrations below c*, the structure of the wire-like aggregates appeared as frozen. It is interesting to note here that the studies as a function of dialysis time and NP concentration exhibited comparable results in terms of morphology diagram. The same sequence of aggregates was obtained with the two procedures. This later experiment indicates however that the structures in regime II are different, one being flexible and stretchable under the application of a field, the other remaining frozen.

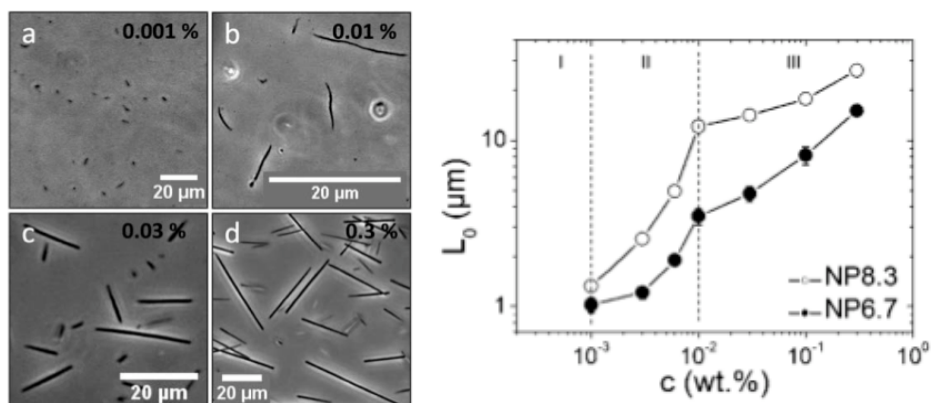

*Figure 7 :* Phase-contrast optical microscopy images (40×) of the NP/polymer dispersions corresponding to the different initial weight concentrations : $10^{-3}$ wt. % (a), $10^{-2}$ wt. % (b), $3 \times 10^{-2}$ wt. % (c) and $3 \times 10^{-1}$ wt. % (d). The data are for dispersion NP6.7 at c = 0.1 wt. % and dialysis time 60 mn.
*Figure 8 :* Average rod length as a function of the initial weight concentration for the NP6.7 (closed symbols) and NP8.3 (open symbols) dispersions. Three morphology regimes including dispersed clusters (Regime I), wire-like aggregates (Regime II) and rigid rods (Regime III) were identified for the nanoparticles co-assembly.

**3.4 – Growth mechanisms for the superparamagnetic nanorods**
Before going into the details of the mechanism, it is important to recall here that the features of the transition observed as a function of the ionic strength (such as in Fig. 5b) are in good agreement with those of the *desalting transition* described in Refs. [27,38]. The desalting transition was identified in the formation of spherical NP clusters using the same cationic $PTEA_{11K}$-*b*-$PAM_{30K}$ copolymers and the same dialysis technique. Moreover, the results for the spherical clusters were obtained with $\gamma$-$Fe_2O_3$ NPs in the absence of a field,[27] and with non magnetic cerium oxide NPs.[38] In these examples, it was found that a critical ionic strength noted $I_S^c$ (around 0.4 M) existed and that above this value the particles remained disperse and unassociated. The desalting transition resembles that disclosed by Stoll and coworkers on the adsorption behavior of a polyelectrolyte chain on an oppositely charged sphere.[39,40] Using Monte Carlo simulations, these authors found that the proportion of monomers adsorbed on a sphere through electrostatic interactions exhibited a steep increase around $I_S = 0.5$ M, indicating a transition between a desorbed and an adsorbed state for the polymers. Similarly, in the desalting process discussed here, we anticipate that the driving force for the association is related to the desorption-adsorption transition of the polyelectrolyte blocks onto the oppositely charged NPs. In the present colloidal system however, the polymers not only adsorb on the particles but also induce their clustering.[38]

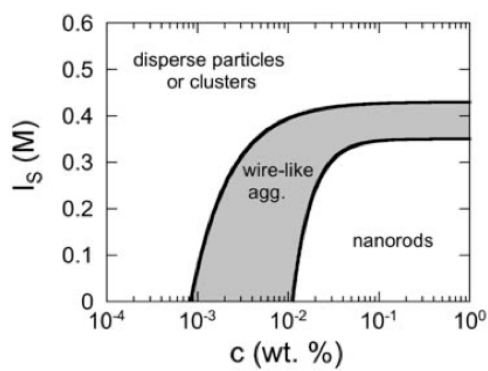

*Figure 9 :* *Morphology diagram for the superparamagnetic aggregates built by the dialysis desalting protocol.*

In order to provide a semi-quantitative description of the rod growth, comments on the magnetic particles and interactions are necessary. The initial magnetic cores of the γ-$Fe_2O_3$ nanoparticles being 6.7 or 8.3 nm in diameter, the magnetic dipolar interactions between particles remain much lower than the thermal energy. In other terms, the dipolar interaction parameter :

$$\lambda_0 = \mu_0 \pi m_S^2 D^3 / 144 k_B T \qquad (2)$$

remains lower than 1, and at the concentration investigated here (c < 0.3 wt. %), the particles do not aligned spontaneously under field.[41, 42] Here, $m_s$ is the magnetization of iron oxide, $\mu_0$ the permeability of vacuum, $k_B$ the Boltzmann constant and T the absolute temperature (T = 298 K). These energetic considerations suggest that the precursors of the rods are not chains of single nanoparticles. We will see below that these precursors are indeed aggregates of particles.

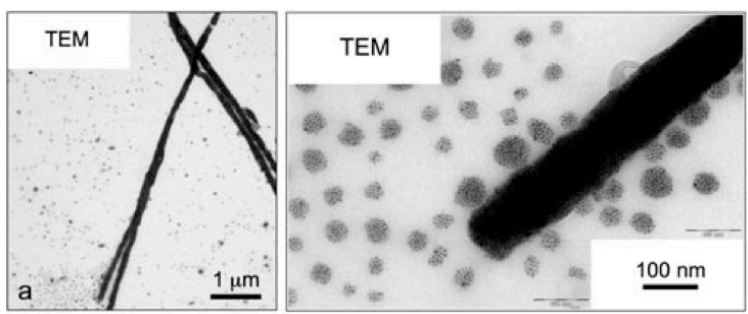

*Figure 10 :* *For rod dispersions that were not washed before the experiment, TEM images display the coexistence of rods and clusters. In this assay, the rods have diameters of 130 ± 10 nm and the clusters are given with a median diameter of 40 nm and a polydispersity of 0.20. Note the remarkable sphericity of the clusters in (b). The magnifications are 6000× and 120000× for a) and b) respectively.*

The TEM images of rods shown in Fig. 3 were recorded after thorough washing of the dispersions. This washing was carried out by magnetophoresis to separate unreacted nanoparticles and submicrometric clusters from the rods. For samples that were not washed before being deposited on the TEM grid, images such as those displayed in Figs. 10a and 10b were obtained. In such cases, rods were found to be coexisting with NP clusters. In Figs. 10, the rods had diameters of 130 ± 10 nm, and the clusters were characterized by a median diameter of 40 nm and a polydispersity of 0.20. Note the remarkable sphericity of the clusters in Fig. 10b and that some aggregates seem to be partially melted onto the rod surface.

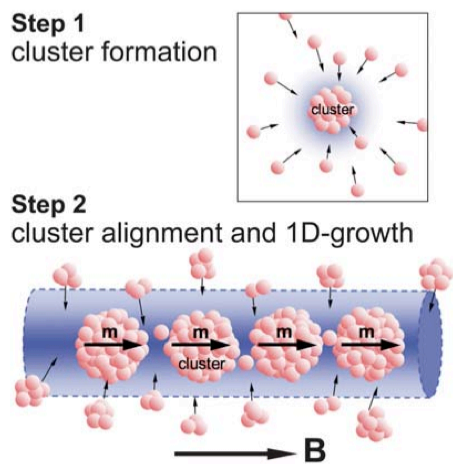

*Figure 11 :* *Schematic representation of the mechanism of rod formation. It consists first in the growth of spherical clusters of particles due to electrostatic complexation between the oppositely charged polymers and particles. In the representation, the particles are negatively charged. The cationic-neutral diblock copolymers were omitted for sake of clarity. The second process is the alignment of the clusters induced by the magnetic dipolar interactions. Concerning the kinetics, the clusters growth and their alignment occur concomitantly, leading to a continuous accretion of particles and clusters on the rodlike structure.*

The comment on the magnetic dipolar interactions and the findings of Fig. 10 suggest that in the dialysis process, even in the presence of a field, there is first the formation of the NP clusters. These clusters are spherical and grow according to a kinetics that is slow at the scale of their diffusion. They are considered as the building blocks or the precursors in the formation of the rods. In absence of a field, their net magnetic moment is zero. This is due to the fact that the reorientations of the single magnetic moments are dominated by the Néel relaxations. Under the application of a field however, the clusters acquire a net magnetic moment which increases linearly with the number n of particles inside a cluster. By analogy with the calculation for single nanoparticles (Eq. 2), it is possible to compute the dipolar interaction parameter $\lambda_C$ between clusters of size $D_C$. $\lambda_C$ reads :

$$\lambda_C = n \phi \mathcal{L}^2 \lambda_0 \qquad (3)$$

where $\phi$ is the volume fraction of nanoparticles in the cluster, $\mathcal{L}$ the value of the Langevin function calculated for the single particles at the applied field, and $\lambda_0$ the expression derived in Eq. 2. n and $\phi$ (= 0.3)[28] are geometrically related to $D_C$ through :[35]

$$D_C = (n/\phi)^{1/3} D \tag{4}$$

From Eqs. 3 and 4, it can be deduced that the interaction parameter between clusters increases with the size as $\lambda_C \sim D_C^3$. Assuming that the chaining between clusters occurs for $\lambda_C > 3$,[43] we found that clusters should align spontaneously under field (here 0.3 T) for $D_C > 50$ nm, corresponding to aggregation numbers larger than $\sim 70$. We propose that the alignment of NP clusters is the second important process of the rod formation. In this respect, the spherical aggregates of Fig. 10 can be viewed as clusters that were not directly incorporated to the rodlike bodies because of their small sizes and magnetic moments. In the process anticipated here, the kinetics of clusters growth and that of their alignment are assumed to be concomitant, so that the accretion of particles or small clusters still continues during the alignment. Fig. 11 illustrates the two mechanisms proposed for the growth kinetics.

## 4 – Conclusion

In this paper, the morphology diagram and the growth mechanisms of nanostructured rods made from single iron oxide nanoparticles were studied. To this aim, a dialysis-based desalting technique was systematically applied to high ionic strength dispersions containing oppositely charged polymers and particles. The major findings of the paper are summarized by Figs. 9 and 11. The role of the particle size and concentration, as well as that of the ionic strength of the dispersions were examined. By tuning the ionic strength, the morphology diagram of the aggregates was found to contain three different domains : single particles (regime I), wire-like filaments (regime II) and rigid rods (regime III). Supracolloidal aggregates could only be observed for ionic strengths below a critical value $I_S^c = 0.42$ M. In Regime II, the precursors of the rigid rods were identified as loose, flexible and stretchable wire-like aggregates. Our experiments show that the growth mechanism depends not only on the ionic strength but also on the initial weight concentration. The concentration diagram could also be divided in three distinct regimes equivalent to that of the ionic strength diagram. We found that higher concentrations favor the formation of long and rigid aggregates. We finally proposed a mechanism for the rod formation where two mechanisms, the formation of spherical clusters and their subsequent alignment by the magnetic field occurred concomitantly, resulting in the formation of highly anisotropic and cylindrical aggregates.

**Acknowledgements**

We thank Olivier Sandre, Régine Perzynski, Jean-Paul Chapel for numerous and fruitful discussions during the course of this work. The Laboratoire Physico-chimie des Electrolytes, Colloïdes et Sciences Analytiques (PECSA, Université Pierre et Marie Curie, Paris, France) is acknowledged for providing us with the nanoparticle dispersions. We are also grateful to Benoit Ladoux from the Laboratoire Matière et Systèmes Complexes (Université Paris-Denis Diderot) for access to microscopy and imaging facility. Aude Michel (PECSA, Université Pierre et Marie Curie, Paris, France) is kindly acknowledged for the TEM experiments. This research was supported in part by Rhodia (France), by the Agence Nationale de la Recherche under the contract BLAN07-3_206866, by the European Community through the project : "NANO3T—

# References

1. N. S. Andrew, K. Eugenii and W. Itamar, *ChemPhysChem*, 2000, **1**, 18.
2. O. D. Velev and E. W. Kaler, *Adv. Mater.*, 2000, **12**, 531.
3. D. J. Norris and A. V. Yu, *Adv. Mater.*, 2001, **13**, 371.
4. A. Islamshah, M. R. Adam, A. N. Larry and K. T. Raymond, *Appl. Phys. Lett*, 2002, **80**, 2761.
5. S. J. Park, T. A. Taton and C. A. Mirkin, *Science*, 2002, **295**, 1503-1506.
6. O. D. Velev and E. W. Kaler, *Langmuir*, 1999, **15**, 3693.
7. C. Goubault, P. Jop, M. Fermigier, J. Baudry, E. Bertrand and J. Bibette, *Physical Review Letters*, 2003, **91**, 260802.
8. L. Y. Wang, X. J. Shi, N. N. Kariuki, M. Schadt, G. R. Wang, Q. Rendeng, J. Choi, J. Luo, S. Lu and C. J. Zhong, *Journal of the American Chemical Society*, 2007, **129**, 2161-2170.
9. Y. Ofir, B. Samanta and V. M. Rotello, *Chemical Society Reviews*, 2008, **37**, 1814-1823.
10. S. A. Corr, S. J. Byrne, R. Tekoriute, C. J. Meledandri, D. F. Brougham, M. Lynch, C. Kerskens, L. O'Dwyer and Y. K. Gun'ko, *J. Am. Chem. Soc.*, 2008, **130**, 4214 - 4215.
11. A. O. Fung, V. Kapadia, E. Pierstorff, D. Ho and Y. Chen, *J. Phys. Chem. C*, 2008, **112**, 15085.
12. J.-H. Park, G. v. Maltzahn, L. Zhang, Michael P. Schwartz, E. Ruoslahti, S. N. Bhatia and M. J. Sailor, *Advanced Materials*, 2008, **20**, 1630–1635.
13. E. M. Furst and A. P. Gast, *Physical Review E*, 2000, **61**, 6732.
14. A. Cebers and I. Javaitis, *Physical Review E*, 2004, **69**, 021404.
15. L. Cohen-Tannoudji, E. Bertrand, L. Bressy, C. Goubault, J. Baudry, J. Klein, J. F. Joanny and J. Bibette, *Physical Review Letters*, 2005, **94**.
16. H. Singh, P. E. Laibinis and T. A. Hatton, *Langmuir*, 2005, **21**, 11500 - 11509.
17. B. A. Evans, A. R. Shields, R. L. Carroll, S. Washburn, M. R. Falvo and R. Superfine, *Nano Lett.*, 2007, **7**, 1428 - 1434.
18. Q. A. Pankhurst, J. Connolly, S. K. Jones and J. Dobson, *J. Phys. D: Appl. Phys.*, 2003, **36**, R167 – R181.
19. J. H. E. Promislow, A. P. Gast and M. Fermigier, *Journal of Chemical Physics*, 1995, **102**, 5492-5498.
20. E. M. Furst, C. Suzuki, M. Fermigier and A. P. Gast, *Langmuir*, 1998, **14**, 7334.
21. H. Singh, P. E. Laibinis and T. A. Hatton, *Nano Letters*, 2005, **5**, 2149-2154.
22. F. Martinez-Pedrero, M. Tirado-Miranda, A. Schmitt and J. Callejas-Fernandez, *Physical Review E*, 2007, **76**, 011405.
23. P. Tierno, T. H. Johansen and T. M. Fischer, *Journal of Physical Chemistry B*, 2007, **111**, 3077-3080.
24. F. E. Osterloh, H. Hiramatsu, R. K. Dumas and K. Liu, *Langmuir*, 2005, **21**, 9709 - 9713.
25. R. Sheparovych, Y. Sahoo, M. Motornov, S. Wang, H. Luo, P. N. Prasad, I. Sokolov and S. Minko, *Chem. Mater.*, 2006, **18**, 591.
26. B. Y. Geng, J. Z. Ma, X. W. Liu, Q. B. Du, M. G. Kong and L. D. Zhang, *Appl. Phys. Lett*, 2007, **90**, 043120.
27. J. Fresnais, J.-F. Berret, B. Frka-Petesic, O. Sandre and R. Perzynski, *Adv. Mater.*, 2008, **20**, 3877-3881.
28. J. Fresnais, J.-F. Berret, L. Qi, J.-P. Chapel, J.-C. Castaing, O. Sandre, B. Frka-Petesic, R. Perzynski, J. Oberdisse and F. Cousin, *Phys. Rev. E*, 2008, **78**, 040401.

29. M. Tanase, L. A. Bauer, A. Hultgren, D. M. Silevitch, L. Sun, D. H. Reich, P. C. Searson and G. J. Meyer, *Nano Letters*, 2001, **1**, 155-158.
30. A. Anguelouch, R. L. Leheny and D. H. Reich, *Applied Physics Letters*, 2006, **89**, 111914.
31. A. O. Fung, V. Kapadia, E. Pierstorff, D. Ho and Y. Chen, *The Journal of Physical Chemistry C*, 2008, **112**, 15085-15088.
32. J. Fresnais, J.-F. Berret, B. Frka-Petesic, O. Sandre and R. Perzynski, *Journal of Physics: Condensed Matter*, 2008, 494216.
33. M. Destarac, W. Bzducha, D. Taton, I. Gauthier-Gillaizeau and S. Z. Zard, *Macromol. Rapid Commun.*, 2002, **23**, 1049 - 1054.
34. M. Jacquin, P. Muller, R. Talingting-Pabalan, H. Cottet, J.-F. Berret, T. Futterer and O. Theodoly, *J. Colloid Interface Sci.*, 2007, **316**, 897-911.
35. J.-F. Berret, *J. Chem. Phys.*, 2005, **123**, 164703.
36. R. Massart, E. Dubois, V. Cabuil and E. Hasmonay, *J. Magn. Magn. Mat.*, 1995, **149**, 1 - 5.
37. A. Sehgal, Y. Lalatonne, J.-F. Berret and M. Morvan, *Langmuir*, 2005, **21**, 9359-9364.
38. J. Fresnais, C. Lavelle and J. F. Berret, *The Journal of Physical Chemistry C*, 2009, **113**, 16371-16379.
39. P. Chodanowski and S. Stoll, *J. Chem. Phys*, 2001, **115**, 4951 - 4960.
40. P. Chodanowski and S. Stoll, *Macromolecules*, 2001, **34**, 2320 - 2328.
41. R. Rosensweig, *Ferrohydrodynamics*, Cambridge University Press, Cambridge, 1985.
42. K. Butter, P. H. H. Bomans, P. M. Frederik, G. J. Vroege and A. P. Philipse, *Nat. Mater*, 2003, **2**, 88.
43. C. Holm and J. J. Weis, *Curr. Opin. Colloid Interface Sci.*, 2005, **10**, 133-140.
44. J. C. Bacri, R. Perzynski, D. Salin, V. Cabuil and R. Massart, *Journal of Magnetism and Magnetic Materials*, 1986, **62**, 36-46.